\documentclass[longauth]{aa}

\usepackage{amsmath}   
\usepackage{amssymb}   
\usepackage{graphicx}  
\usepackage{booktabs}  
\usepackage{xspace}    

\usepackage{orcidlink}

\usepackage{hyperref}  

\definecolor{myblue}{HTML}{1F77B4}
\definecolor{mygreen}{HTML}{2CA02C}
\definecolor{myred}{HTML}{D62728}
\definecolor{mymagenta}{HTML}{D33682}
\definecolor{codepurple}{HTML}{C42043}

\hypersetup{
	unicode=true,           
	colorlinks=true,        
	linkcolor=myred,        
	citecolor=myblue,       
	filecolor=mymagenta,    
	urlcolor=mygreen        
}

\usepackage{longtable}

\usepackage{subcaption} 

\usepackage{silence}
\usepackage{silence}
\newcommand{\halpha}{H\,\textsc{$\alpha$}\xspace}

\newcommand{\OIdoublet}{[O\,\textsc{i}]~$\lambda\lambda$6300, 6364\xspace} 
\newcommand{\OIsinglet}{[O\,\textsc{i}]~$\lambda$5577\xspace} 
\newcommand{\NIIdoublet}{[N\,\textsc{ii}]~$\lambda\lambda$6548, 6583\xspace} 
\newcommand{\NIIsinglet}{[N\,\textsc{ii}]~$\lambda$5754\xspace} 
 
\newcommand{\NIIdiag}{$f_{[\text{N}\,\textsc{ii}]}$\xspace}
\newcommand{\el}[2]{$^{#1}$#2\xspace}
\newcommand{\Ni}{\el{56}{Ni}}

\newcommand{\kms}{\text{~km~s$^{-1}$}\xspace}
\newcommand{\msun}{\text{M$_{\odot}$}\xspace}

\newcommand{\msunperyr}{\text{M$_{\odot}$~yr$^{-1}$}\xspace}

\newcommand{\mop}{SN~2022mop\xspace}
\newcommand{\pvb}{SN~2020pvb\xspace}
\newcommand{\ip}{SN~2009ip\xspace}
\newcommand{\fyq}{SN~2023fyq\xspace}
\newcommand{\jbu}{SN~2016jbu\xspace}

\newcommand{\bh}{SN~2015bh\xspace}
\newcommand{\host}{IC~1496\xspace}
\newcommand{\PS}{Pan-STARRS\xspace}

\newcommand{\etacar}{Eta~Carinae\xspace}
\newcommand{\tzo}{T\.ZO\xspace}

\newcommand{\autophot}{\texttt{AutoPhOT}\xspace}
\newcommand{\sumo}{\texttt{SUMO}\xspace}

\usepackage{hyperref}

\definecolor{xlinkcolor}{rgb}{0,0,1}  
\hypersetup{
	colorlinks=true,        
	linkcolor=xlinkcolor,   
	citecolor=xlinkcolor,   
	urlcolor=xlinkcolor     
}

\begin{document}
	
	\title{ \vspace{-0.8cm} Precursor Activity Preceding Interacting Supernovae -- I: Bridging the Gap with \mop}
	\titlerunning{The case of \mop}
	
	\author{S. J. Brennan \thanks{Email: \href{mailto:sean.brennan@astro.su.se}{sean.brennan@astro.su.se}} \fnmsep \inst{1}\href{https://orcid.org/0000-0003-1325-6235}{\includegraphics[scale=0.05]{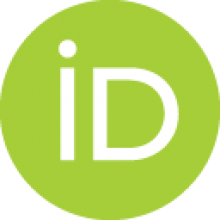}}
\and
S. Barmentloo\inst{1}\href{https://orcid.org/0000-0003-4800-2737}{\includegraphics[scale=0.05]{orcid-ID.png}}
\and
S. Schulze\inst{2}\href{https://orcid.org/0000-0001-6797-1889}{\includegraphics[scale=0.05]{orcid-ID.png}}
\and
K. W. Smith\inst{3,4}\href{https://orcid.org/0000-0001-9535-3199}{\includegraphics[scale=0.05]{orcid-ID.png}}
\and
R. Hirai\inst{5,6,7}\href{https://orcid.org/0000-0002-8032-8174}{\includegraphics[scale=0.05]{orcid-ID.png}}
\and
J. J. Eldridge\inst{8}\href{https://orcid.org/0000-0002-1722-6343}{\includegraphics[scale=0.05]{orcid-ID.png}}
\and
M. Fraser\inst{9}\href{https://orcid.org/0000-0003-2191-1674}{\includegraphics[scale=0.05]{orcid-ID.png}}
\and
H. F. Stevance\inst{3}\href{https://orcid.org/0000-0002-0504-4323}{\includegraphics[scale=0.05]{orcid-ID.png}}
\and
S. J. Smartt\inst{3,4}\href{https://orcid.org/0000-0002-8229-1731}{\includegraphics[scale=0.05]{orcid-ID.png}}
\and
S. Anand\inst{10}\href{https://orcid.org/0000-0003-3768-7515}{\includegraphics[scale=0.05]{orcid-ID.png}}
\and
A. Aryan\inst{11}\href{https://orcid.org/0000-0002-9928-0369}{\includegraphics[scale=0.05]{orcid-ID.png}}
\and
T.-W. Chen\inst{11}\href{https://orcid.org/0000-0002-1066-6098}{\includegraphics[scale=0.05]{orcid-ID.png}}
\and
K. K. Das\inst{10}\href{https://orcid.org/0000-0001-8372-997X}{\includegraphics[scale=0.05]{orcid-ID.png}}
\and
A. J. Drake\inst{12}\href{https://orcid.org/0000-0000-0000-0000}{\includegraphics[scale=0.05]{orcid-ID.png}}
\and
C. Fransson\inst{1}\href{https://orcid.org/0000-0001-8532-3594}{\includegraphics[scale=0.05]{orcid-ID.png}}
\and
A. Gangopadhyay\inst{1}\href{https://orcid.org/0000-0002-3884-5637}{\includegraphics[scale=0.05]{orcid-ID.png}}
\and
A. Gkini\inst{1}\href{https://orcid.org/0009-0000-9383-2305}{\includegraphics[scale=0.05]{orcid-ID.png}}
\and
W. V. Jacobson-Gal{\'a}n\thanks{NASA Hubble Fellow}\fnmsep\inst{10}\href{https://orcid.org/0000-0002-3934-2644}{\includegraphics[scale=0.05]{orcid-ID.png}}
\and
A. Jerkstrand\inst{1}\href{https://orcid.org/0000-0001-8005-4030}{\includegraphics[scale=0.05]{orcid-ID.png}}
\and
J. Johansson\inst{13}\href{https://orcid.org/0000-0001-5975-290X}{\includegraphics[scale=0.05]{orcid-ID.png}}
\and
M. Nicholl\inst{14}\href{https://orcid.org/0000-0002-2555-3192}{\includegraphics[scale=0.05]{orcid-ID.png}}
\and
G. Pignata\inst{15}\href{https://orcid.org/0000-0003-0006-0188}{\includegraphics[scale=0.05]{orcid-ID.png}}
\and
N. Sarin\inst{1,16}\href{https://orcid.org/0000-0003-2700-1030}{\includegraphics[scale=0.05]{orcid-ID.png}}
\and
A. Singh\inst{1}\href{https://orcid.org/0000-0003-2091-622X}{\includegraphics[scale=0.05]{orcid-ID.png}}
\and
J. Sollerman\inst{1}\href{https://orcid.org/0000-0003-1546-6615}{\includegraphics[scale=0.05]{orcid-ID.png}}
\and
S. Srivastav\inst{3}\href{https://orcid.org/0000-0003-4524-6883}{\includegraphics[scale=0.05]{orcid-ID.png}}
\and
B.F.A. van Baal\inst{1}\href{https://orcid.org/0009-0001-3767-942X}{\includegraphics[scale=0.05]{orcid-ID.png}}
\and
K. C. Chambers\inst{17}\href{https://orcid.org/0000-0001-6965-7789}{\includegraphics[scale=0.05]{orcid-ID.png}}
\and
M. W. Coughlin\inst{18}\href{https://orcid.org/0000-0002-8262-2924}{\includegraphics[scale=0.05]{orcid-ID.png}}
\and
H. Gao\inst{17}\href{https://orcid.org/0000-0003-1015-5367}{\includegraphics[scale=0.05]{orcid-ID.png}}
\and
M. J. Graham\inst{10}\href{https://orcid.org/0000-0002-3168-0139}{\includegraphics[scale=0.05]{orcid-ID.png}}
\and
M. E. Huber\inst{17}\href{https://orcid.org/0000-0003-1059-9603}{\includegraphics[scale=0.05]{orcid-ID.png}}
\and
C.-C. Lin\inst{17}\href{https://orcid.org/0000-0002-7272-5129}{\includegraphics[scale=0.05]{orcid-ID.png}}
\and
T. B. Lowe\inst{19}\href{https://orcid.org/0000-0002-9438-3617}{\includegraphics[scale=0.05]{orcid-ID.png}}
\and
E. A. Magnier\inst{17}\href{https://orcid.org/0000-0002-7965-2815}{\includegraphics[scale=0.05]{orcid-ID.png}}
\and
F. J. Masci\inst{20}\href{https://orcid.org/0000-0002-8532-9395}{\includegraphics[scale=0.05]{orcid-ID.png}}
\and
J. Purdum\inst{12}\href{https://orcid.org/0000-0003-1227-3738}{\includegraphics[scale=0.05]{orcid-ID.png}}
\and
A. Rest\inst{21,22}\href{https://orcid.org/0000-0002-4410-5387}{\includegraphics[scale=0.05]{orcid-ID.png}}
\and
B. Rusholme\inst{20}\href{https://orcid.org/0000-0001-7648-4142}{\includegraphics[scale=0.05]{orcid-ID.png}}
\and
R. Smith\inst{12}\href{https://orcid.org/0000-0001-7062-9726}{\includegraphics[scale=0.05]{orcid-ID.png}}
\and
I. A. Smith\inst{23}\href{https://orcid.org/0000-0001-8605-5608}{\includegraphics[scale=0.05]{orcid-ID.png}}
\and
J. W. Tweddle\inst{3}\href{https://orcid.org/0009-0004-5681-545X}{\includegraphics[scale=0.05]{orcid-ID.png}}
\and
R.J. Wainscoat\inst{17}\href{https://orcid.org/0000-0002-1341-0952}{\includegraphics[scale=0.05]{orcid-ID.png}}
\and
T. de Boer\inst{17}\href{https://orcid.org/0000-0001-5486-2747}{\includegraphics[scale=0.05]{orcid-ID.png}}
} 


	\institute{The Oskar Klein Centre, Department of Astronomy, Stockholm University, AlbaNova, SE-10691 Stockholm, Sweden
\and
Center for Interdisciplinary Exploration and Research in Astrophysics, Northwestern University, 1800 Sherman Ave., Evanston, IL 60201, USA
\and
Department of Physics, University of Oxford, Denys Wilkinson Building, Keble Road, Oxford OX1 3RH, UK
\and
Astrophysics Research Centre, School of Mathematics and Physics, Queen’s University Belfast, BT7 1NN, UK
\and
Astrophysical Big Bang Laboratory, Cluster for Pioneering Research, RIKEN, Wako, Saitama 351-0198, Japan
\and
School of Physics and Astronomy, Monash University, Clayton, VIC 3800, Australia
\and
OzGrav: The ARC Centre of Excellence for Gravitational Wave Discovery, Clayton, VIC 3800, Australia
\and
Department of Physics, University of Auckland, Private Bag 92019, Auckland, New Zealand
\and
School of Physics, University College Dublin, L.M.I. Main Building, Beech Hill Road, Dublin 4 D04 P7W1, Ireland
\and
Division of Physics, Mathematics and Astronomy, California Institute of Technology, Pasadena, CA 91125, USA
\and
Graduate Institute of Astronomy, National Central University, 300 Jhongda Road, 32001 Jhongli, Taiwan
\and
Caltech Optical Observatories, California Institute of Technology, 1200 E. California Blvd., Pasadena, CA  91125, USA
\and
The Oskar Klein Centre, Department of Physics, Stockholm University, AlbaNova, SE-10691 Stockholm, Sweden
\and
Astrophysics Research Centre, School of Mathematics and Physics, Queens University Belfast, Belfast BT7 1NN, UK
\and
Instituto de Alta Investigaci{\'o}n, Universidad de Tarapac{\'a}, Casilla 7D, Arica, Chile
\and
Nordita, Stockholm University and KTH Royal Institute of Technology, Hannes Alfv{\'e}ns v{\"a}g 12, SE-106 91 Stockholm, Sweden
\and
Institute for Astronomy, University of Hawaii, 2680 Woodlawn Drive, Honolulu HI 96822, USA
\and
School of Physics and Astronomy, University of Minnesota, Minneapolis, Minnesota 55455, USA
\and
Institute for Astronomy, University of Hawaii 34 Ohia Ku St., Makawao, HI 96768, USA
\and
IPAC, California Institute of Technology, 1200 E. California Blvd, Pasadena, CA 91125, USA
\and
Space Telescope Science Institute, 3700 San Martin Drive, Baltimore, MD 21218, USA
\and
Department of Physics and Astronomy, The Johns Hopkins University, 3400 North Charles Street, Baltimore, MD 21218, USA
\and
Institute for Astronomy, University of Hawaii, 34 Ohia Ku St., Pukalani, HI 96768-8288, USA
}
	
	
	\abstract{
		Over the past two decades, an increasing number of transients have exhibited luminous activity at their explosion sites weeks to years before an interacting supernova (SN) is observed. For some objects, this pre-SN activity is typically linked to large-scale mass-loss events preceding core collapse, yet its triggering mechanism and the underlying explosion process remain uncertain. We present \mop, which was initially observed in August 2022 exhibiting nebular emission, including [O~{\sc i}], Mg~{\sc i}], and [Ca~{\sc ii}], resembling the late-time ($\gtrsim 200$ days post-explosion) spectrum of a stripped-envelope SN (SESN) from a progenitor with $M_{\text{ZAMS}} \lesssim 18~\msun$. \mop shows strong ($\sim$1~mag) repeating undulations in its light curve, suggesting late-time interaction. In mid-2024, the transient re-brightened for several months before a Type IIn SN ($r_{\text{peak}} = -18.2$~mag) was observed in December 2024, closely resembling the evolution of \ip. By triangulating both transients using \PS images, we determine that both transients are coincident within $\sim$3~pc. Given the environment, the chance alignment of two isolated SNe is unlikely.  We propose a merger-burst scenario -- a compact object formed in 2022, is kicked into an eccentric orbit, interacts with its hydrogen-rich companion over subsequent months, and ultimately merges with its companion, triggering a Type IIn SN-like transient.
	}
	
	\keywords{supernovae: individual (\mop) -- stars: mass loss -- stars: circumstellar matter -- stars: neutron}
	
	\maketitle
	
	\nolinenumbers
	\section{Introduction}\label{sec:introduction}
	
	The liberation of material from the stellar surface plays a crucial role in the life cycle and eventual demise of massive stars \citep[$>$8--10~\msun; ][]{Chiosi1986,Smith2014b,Puls2008,Meynet2015,Leung2021,Laplace2021}.                                                            
	The effects of mass loss are typically observed during the final SN explosions, where the progenitor's chemical composition becomes apparent in the first few days to weeks following core-collapse \citep{Niemela1985,Smartt2009,Gal-Yam2014,Yaron2017,Gal-Yam2017}. Massive stars continuously eject material from their surface via stellar winds \citep{Vink2024} spanning several orders of magnitude ($\sim10^{-8} - 10^{-4}$ \msunperyr), depending on their characteristics (e.g. mass, metallicity, and rotation). However, this may not be sufficient to account for the large amounts of confined material \citep[as much as 0.1~\msun within $10^{15}$ cm; ][]{Tsuna2023,Ransome2024} inferred from observations. The majority of massive stars are found in multi-stellar systems \citep{Sana2012, Offner2023,Marchant2024} and will likely interact with a companion star \citep{Zapartas2017}. Mass transfer can occur, removing large amounts of material via Roche-lobe overflow \citep[RLOF; ][]{deMink2013, Eldridge2017}. The effect of binaries drastically alters the pre-SN progenitor, the amount of  ejected circumstellar material (CSM), and the environment in which it explodes \citep{Laplace2021}, complicating our understanding of their evolution and interaction.
	
	In this work we present \mop, which initially resembled a hydrogen-poor SESN during its nebular phase in August 2022. In late 2024, a transient at the position of \mop began to slowly re-brightened over several months, before an SN-like light curve was observed, with spectral observations indicating a hydrogen-rich interacting Type IIn SN classification. Section \ref{sec:observations} outlines the discovery of \mop in 2022, its properties, as well as the photometric and spectroscopic observations. In Section \ref{sec:methods}, we investigate the alignment of the two events, as well as the environment in which the transient(s) occurred. Section \ref{sec:discussion} provides potential explanations for \mop, including the possibility of two separate SN explosions, as well as the second SN-like event being triggered by the merging of a hydrogen-rich star and a newly formed compact object.
	
	\section{Discovery and Follow-up Observations}\label{sec:observations}
	
	\subsection{\mop and Host Galaxy}\label{ssec:discovery}
	
	\begin{figure}[!ht]
		\centering
		\includegraphics[width=\columnwidth]{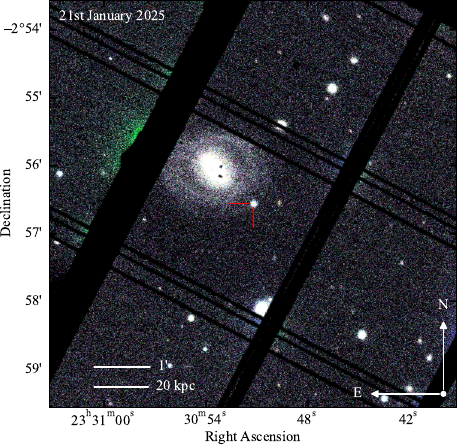} 
		\caption{Image of \mop and \host, composed of 30s \PS $gri$ exposures taken near the peak of the 2024 transient. \mop, marked with red crosshairs, lies in a region of diffuse star formation approximately 16~kpc from the core of \host, located at a luminosity distance of 70.7~Mpc. The diagonal black gaps represent chip gaps on the \PS/GPC1 CCD detector.}
		\label{fig:finder}
	\end{figure}
	
	\mop\footnote{Also classified as ATLAS24saw, Gaia25aaf, GOTO25gj, PS22fhf, and ZTF22aanyxmg.} was discovered by the \PS collaboration on 12 June 2022 \citep{Chambers2022}, at a magnitude of $w \sim 20.45$~mag. The location of the transient is approximately 46$\arcsec$ from the core of the lenticular galaxy, \host, as shown in Fig. \ref{fig:finder}. With a redshift of $z = 0.0169$ \citep{Jones2004, Jones2009}, this corresponds to a luminosity distance of 70.7~Mpc (distance modulus of 34.24~mag) and a separation of 16~kpc between \mop and the centre of \host\footnote{We assume a Hubble constant $H_{0} = 73$~km~s$^{-1}$~Mpc$^{-1}$, $\Omega_{\Lambda} = 0.73$, and $\Omega_{\rm M} = 0.27$ \citep{Spergel2007}.}. We apply a correction for foreground Milky Way (MW) extinction using ${\rm R_V} = 3.1$ and ${\rm E (B - V)_{MW}} = 0.048$~mag \citep{Schlafly2011}, with the extinction law given by \citet{Cardelli1989}. No correction is made for any host extinction, as \mop is located in an isolated region and is unlikely to experience significant dust attenuation, as well as no strong evidence for Na I D absorption \citep{Poznanski2012}.
	
	\subsection{The 2024 Rebrightening of \mop}\label{ssec:rebrightening}
	
	For clarity and consistency with similar objects \citep[e.g. \ip; ][]{Smith2008,Foley2011,Pastorello2013,Margutti2014}, we refer to the entire evolution of the transient as \mop, distinguishing between the 2022 and 2024 transients accordingly. \mop re-emerged in August 2024 and was identified (priv. comm.) as a late-time rebrightening of an SESNe \citep[e.g. ][]{Chen2018, Sollerman2020}. On the 28th December 2024, \mop exhibited a sharp increase in brightness \citep{Srivastav2025}, with a light curve evolution now resembling the pre-SN activity seen in hydrogen-rich Type IIn/Ibn SNe shortly before core collapse \citep{Pastorello2013,Strotjohann2021,Brennan2024}. The 2024 transient peaked in the $r$-band on 13 January 2025 (MJD: 60688.39) with a peak apparent magnitude of $r = 16.40\pm0.02$~mag, with spectral observations showing strong emission from the Balmer series \citep{Sollerman2025}. 
	
	\subsection{Photometric Observations and Evolution}\label{ssec:Photometry}
	
	\begin{figure*}[!ht]
		\centering
		\includegraphics[width=\textwidth]{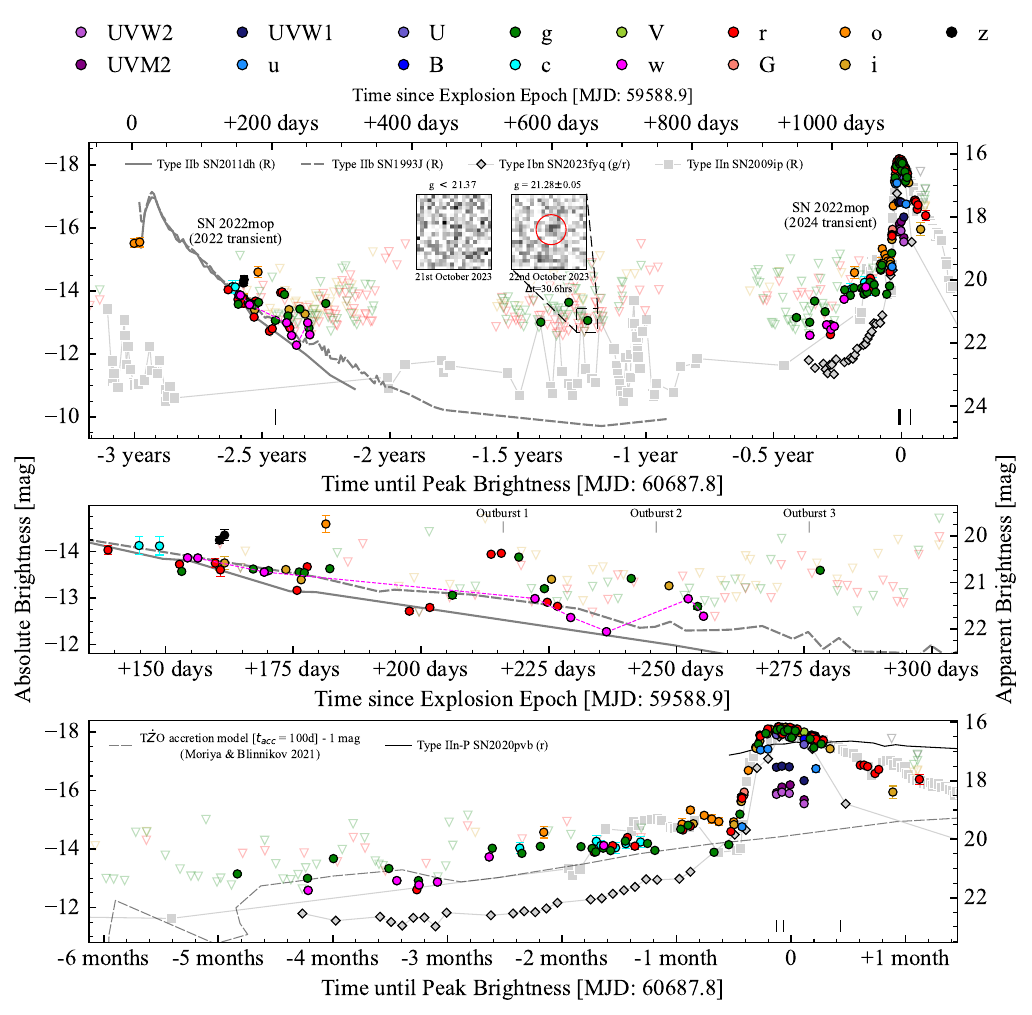 } 
		\caption{\textit{Upper Panel:} Optical light curve evolution of \mop over $\sim3$ years since discovery, covering both the 2022 and 2024 transients. Vertical black lines indicate when spectra were obtained. \textit{Middle panel:} The 2022 event resembles a hydrogen-poor SN (see Section~\ref{sec:modelling}) but exhibits a bumpy decay seen in \PS and ZTF photometry, that is inconsistent with purely \Ni-powered emission. These outburst or flares likely require additional energy input from CSM interaction. \textit{Lower panel:} The 2024 event resembles an interacting hydrogen-rich transient, such as \ip, with a $\sim4$-month-long pre-SN rise, followed by a transient that photometrically and spectroscopically resembles an interacting SN.}
		\label{fig:lightcurve}
	\end{figure*}
	
	Science-ready images obtained as part of the ZTF survey \citep{Bellm2019,Graham2019,Dekany2020,Masci2019} were acquired in $gri$ from the NASA/IPAC Infrared Science Archive (IPAC\footnote{\href{https://irsa.ipac.caltech.edu/applications/ztf/}{https://irsa.ipac.caltech.edu/applications/ztf/}}) service. Template-subtracted photometry was performed using the \autophot\ pipeline \citep{Brennan2022d}; see Appendix~\ref{A:obs_log} for further details. No significant activity or outbursts were detected in pre-discovery images extending back to the beginning of ZTF observations (circa-2018), with a limiting absolute magnitude of approximately $-14$~mag. Photometry from the ATLAS forced-photometry server\footnote{\href{https://fallingstar-data.com/forcedphot/}{https://fallingstar-data.com/forcedphot/}} \citep{Tonry2018,Smith2020,Shingles2021a} was obtained in the $c$ and $o$ bands. We computed the weighted average of the fluxes of observations on a nightly cadence. A quality cut at 5$\sigma$ was applied to the resulting nightly fluxes for both filters, after which the measurements were converted to the AB magnitude system.
	
	The field of \mop was observed as part of regular \PS  \citep{Chambers2016a} operations in 2022 in $w$ and $i$ filters. Following the 2024 rebrightening, we obtained targeted follow-up observations with PS in the $grizy$ filters. The PS data were processed with the Image Processing Pipeline \citep{Magnier2020a, Waters2020,Magnier2020c,Magnier2020b}. We also utilised the 40~cm SLT telescope at the Lulin observatory and obtained images in SDSS $ugriz$ filters as part of the Kinder collaboration \citep[][]{2024arXiv240609270C}. Images were calibrated using a custom built pipeline\footnote{\href{https://hdl.handle.net/11296/98q6x4}{https://hdl.handle.net/11296/98q6x4}} and photometred using the \autophot pipeline. We retrieve Dark Energy Camera (DECam) images covering the site of \mop from the NOIRLAB data archive\footnote{\href{https://datalab.noirlab.edu/sia.php}{https://datalab.noirlab.edu/sia.php}}, including data from 2022\footnote{\href{https://astroarchive.noirlab.edu/portal/search/}{https://astroarchive.noirlab.edu/portal/search/}} (\#2020B-0053, PI. Brout). No activity is detected prior to 2022. \mop is visible in unfiltered images (calibrated to the $r$-band) from the Catalina Sky Survey's \citep{Drake2009} 1.5m Mount Lemmon Survey (MLS) telescope, taken on 27th May 2022, and was observed again on 12th September 2024. No clear prior priot to 2022 is seen in additional MLS observations dating back to September 2010. Additionally, we photometer unfiltered images from Spacewatch \citep{McMillan2015}, also calibrated to the $r$-band, taken sporadically from 2003 to 2015. No significant activity is detected.
	
	The field of \mop was observed with the Wide-field Infrared Survey Explorer (WISE) instrument in the W1 and W2 bands during the NEOWISE-reactivation mission \citep[NEOWISE-R;][]{mainzer2014}. Single exposures in the W1 and W2 bands were collected from the NASA/IPAC Infrared Science Archive (IRSA) and coadded using the ICORE service \citep{masci2013}. Reference images were created by coadding data from 2015, which were subtracted from the subsequent images. 
	
	UV photometry was obtained using the Ultra-Violet Optical Telescope (UVOT) onboard the \textit{Neil Gehrels Swift Observatory} \citep{Gehrels2004,Roming2005}. Image reduction was performed on non-template subtracted images using the \textit{Swift} \texttt{HEAsoft} package\footnote{\href{https://heasarc.gsfc.nasa.gov/docs/software/heasoft/}{https://heasarc.gsfc.nasa.gov/docs/software/heasoft/}}. A log of photometric observations is provided in Table~\ref{tab:SN2022mop_photometry_log} and illustrated in Fig.~\ref{fig:lightcurve}.
	
	\begin{figure*}[!ht]
		\centering
		\includegraphics[width=\textwidth]{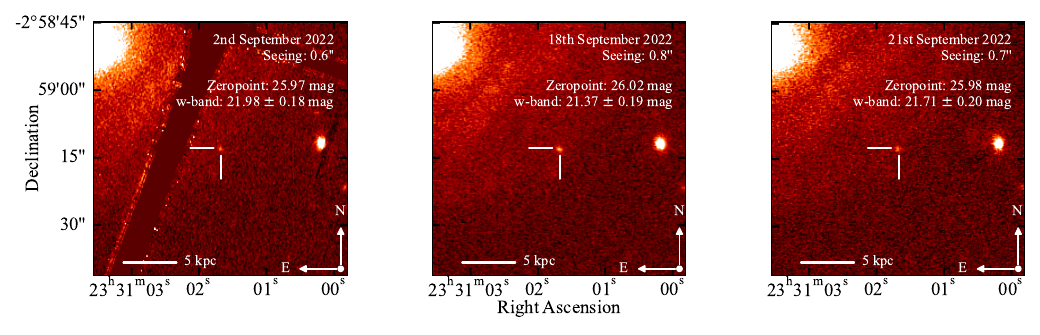 } 
		\caption{\PS $w$-band cutouts of \mop from September 2022 during the transient's nebular phase. Over approximately three weeks, the transient brightened by $\sim$0.6 magnitudes before fading, see Fig.~\ref{fig:lightcurve}. This rebrightening may indicate late-time CSM interaction. See Section~\ref{sec:mergerburst} for further discussion.}
		\label{fig:mosaic}
	\end{figure*}
	
	Our photometric coverage of the 2022 transient is limited due to the field being in solar conjunction when the transient was expected to reach peak brightness. Two ATLAS observations (both S/N $>$ 5) in the $o$-band capture the initial rise of \mop. Following its return from solar conjunction, photometry from \PS, ATLAS, and ZTF captures the nebular phase decline until the transient fades beyond detection limits.
	
\begin{figure}[!ht]
		\centering
		\includegraphics[width=\columnwidth]{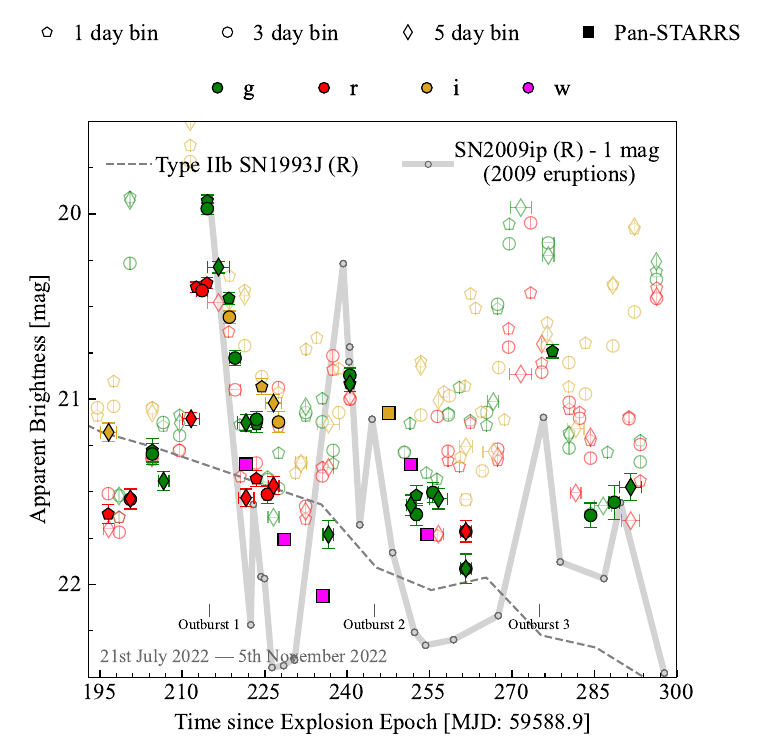 } 
		\caption{Re-binned ZTF light curve of \mop during the 2022 decline. Images are stacked into their respective bins, given in the upper legend, and processed as described in Section~\ref{ssec:Photometry}. Filled markers denote points with S/N $> 3$, where the source meets detection criteria via artificial source injection. The 2022 decline phase exhibits several outbursts/flares, roughly 27 days apart, loosely resembling the 2009 eruptions of \ip \citep{Pastorello2013} (solid grey line, fit by eye). The dashed line the the radioactive powered tail of SN~1993J, highlighting the strong divergence from a \Ni powered decline.}
		\label{fig:binned_lightcurve}
	\end{figure}

	\mop exhibits outburst-like events from $\sim$215 days post-explosion (see Appendix~\ref{A:explosion_epoch} for the explosion time ($t_{\text{exp}}$) estimation). During its smooth decline, a $>1$~mag outburst occurs in all bands. Neighbouring non-detections constrain this rise to at least 1.5 mag in $\sim$3 days, followed by a $\sim$20-day decline. Around 250 days post-explosion, a $\sim$0.6~mag bump is observed (Fig.~\ref{fig:mosaic}), primarily in the $w$-band from \PS, with sporadic detections in ZTF $g$ and $r$. A potential third outburst is observed around $\sim$275 days post-explosion, although the cadence around this epoch is low. These outbursts may be repeating, with a period of approximately 27 days \citep[about twice as long as that reported for SN~2022jli; ][]{Moore2023,Chen2023}, as well as decreasing magnitude. To investigate further, we bin available ZTF $gri$ images during this phase. As shown in Fig. \ref{fig:binned_lightcurve}, there is a resemblance between these outburst and the initial appearance of \ip in 2009, overall resembling a sawtooth like lightcurve with the peaks separated by $\sim$30~days, potentially suggesting a similar powering mechanism. It should be noted that spectra of \ip around this time show strong sings of interaction with H-rich material \citep{Smith2010,Foley2011,Pastorello2013}, unlike what is observed for \mop. Further efforts are required to rigorously account for this flaring, which is beyond the immediate goals of this work. 
	
	\mop was detected again around 1.5 years post-explosion in the $g$-band that again, must be occurring on rapid timescales (i.e. a few a days ), as indicated by neighbouring non-detections. This behaviour is reminiscent of the 2010 -- 2011 eruptions from the progenitor of \ip \citep{Pastorello2013}, reaching similar magnitudes and timescales. While our optical detections are this phase are limited, we also detect \mop in the mid-infrared, see Fig. \ref{fig:wise_lightcurve}. The origin of this emission remains unclear, though it has been proposed that such pre-SN outbursts are driven by ejecta or outflow interaction with clumpy CSM or by binary interaction \citep{Soker2013,Kashi2013}. 
	
	Following its 2024 solar conjunction, \mop was observed to brighten continuously\footnote{\mop may be experiencing flickering events during this rise, but due to the low cadence observations this is difficult to falsify.} from August 2024 over the subsequent four months, rising by approximately one magnitude in all bands. Although the detection cadence is suboptimal, the light curve shows some structure, beginning with a bump lasting about one month, followed by a more gradual rise. Similar to \ip's Event A, which exhibited a decline of approximately one magnitude before the onset of Event B, \mop shows a similar trend, mainly seen in the $g$-band. Subsequently, \mop exhibits a SN-like rise on 28th December 2024, with a rise time of $\sim12$~days, typical of Type IIn SNe \citep{Taddia2013,Nyholm2020}.
	
	As the 2024 transient evolves post-peak, the field begins to set, limiting observations. Several epochs were obtained using ZTF and Lulin facilities around 20 days post-peak. A single ZTF $r$-band data point shows a small bump in the light curve at approximately 25 days post-peak. A similar bump was observed for \ip \citep[e.g. ][]{Kashi2013} and is common in the post-peak evolution of several interacting transients \citep{Nyholm2020}. 
	
	\subsection{Spectroscopic Observations and Appearance}\label{ssec:Spectroscopy}
	
	\begin{figure*}[!ht]
		\centering
		\includegraphics[width=\textwidth]{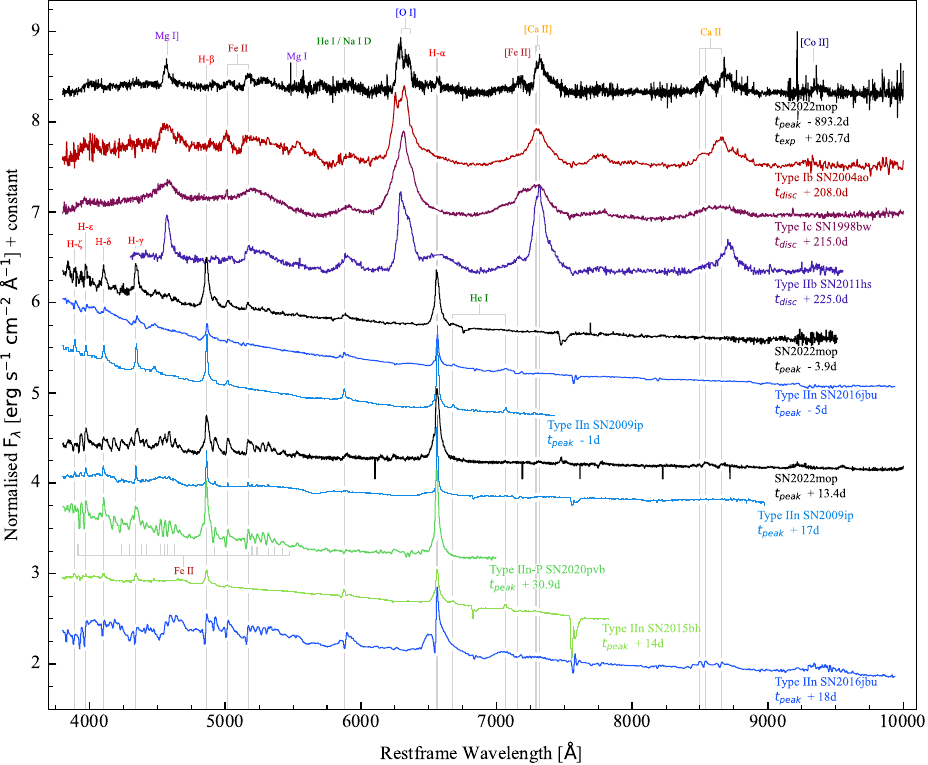 } 
		\caption{Spectral comparison of \mop with other relevant SNe. All spectra have been corrected for redshift and MW extinction. \mop was observed in August 2022 and resembled a SESN during the nebular phase, e.g. SNe 2004ao \citep{Modjaz2014}, 1998bw \citep{Patat2001}, and 2011hs \citep{Bufano2014}. Over two years later, \mop experienced a dramatic rebrightening, resembling a hydrogen-rich interacting transient similar to \ip \citep{Pastorello2013,Fraser2013} and \bh \citep{Elias-Rosa2016,Thone2017}. Post-peak spectral observations closely match those of interacting transients that show pre-SN activity including \jbu \citep{Kilpatrick2018,Brennan2022c}, and classified Type IIn-P SNe \citep[\pvb; ][]{Elias-Rosa2024}.}
		\label{fig:compare_spectra}
	\end{figure*}
	
	\mop was observed on 1st August 2022 as part of the Census of the Local Universe (CLU) survey \citep{Cook2019,De2020} using the Low Resolution Imaging Spectrometer \citep[LRIS; ][]{Oke1995} on the 10-m Keck telescope; see Fig. \ref{fig:compare_spectra}. Prior to these observations, \mop was in solar conjunction meaning the peak lightcurve phase was missed. Due to strong emission from [O {\sc i}], Mg {\sc i}], and [Ca {\sc ii}], \mop resembled a typical nebular SESN and received minimal follow-up. 
	
	Motivated by the 2024 rebrightening in December 2024, follow-up spectra were obtained using the Nordic Optical Telescope (NOT) with the Alhambra Faint Object Spectrograph and Camera (ALFOSC) and Keck/LRIS. The spectra were reduced in a standard manner using  \texttt{PypeIt} \citep{pypeit:zenodo,pypeit:joss_arXiv,pypeit:joss_pub} and \texttt{LPipe} \citep{Perley2019} respectively, with observations coordinated using the FRITZ data platform \citep{Walt2019,Coughlin2023}. Further discussion of spectroscopic features is given in Section. \ref{sec:modelling} and \ref{sec:09ip}. An observation log of spectral observations is given in Table. \ref{tab:SN2022mop_spectroscopy_log} and illustrated in Fig. \ref{fig:compare_spectra} and \ref{fig:all_spectra}.
	
	\section{Methods and Analysis}\label{sec:methods}
	
	\subsection{Coincidence Between the 2022 and 2024 Transients}\label{sec:coincidence}
	
	\begin{figure}[!ht]
		\centering
		\includegraphics[width=\columnwidth,scale = 0.75]{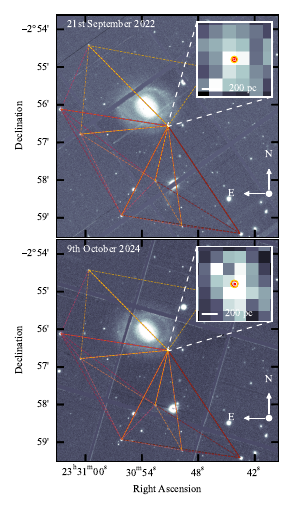 } 
		\caption{Triangulation of the 2024 source based on the fitted position of the 2022 source using triangulation of matching sources both images. Positions were determined using a point-spread functions built using bright, isolated sources in each image. Each inset shows the fitted position (gold) and the triangulation contours (1 and 2$\sigma$ regions given in red) for each epoch.}
		\label{fig:triangulation}
	\end{figure}
	
	\begin{figure}[!ht]
		\centering
		\includegraphics[width=\columnwidth]{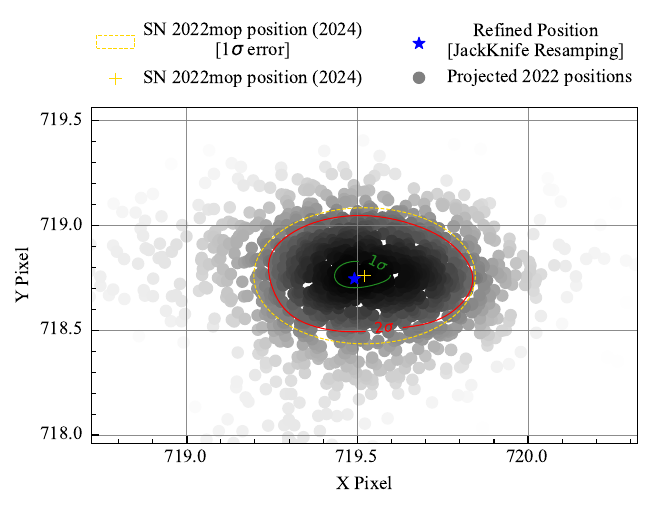} 
		\caption{The 2022 transient positions (black points) projected onto the 2024 image using \texttt{TransientTriangulator}. The gold region represents the 1$\sigma$ fitted position of the 2024 transient. The alignment of the 2022 and 2024 transients within their 1$\sigma$ uncertainties results in an offset of 0.020 $\pm$ 0.25 pixels or $\sim$2.3 parsecs.}
		\label{fig:contours}
	\end{figure}
	
	We investigate the possibility that the 2022 and 2024 transient are two isolated events that coincidentally occupy the same pixel on the CCD detector \citep[similar to the recent SN~2020acct; ][]{Angus2024}. Using common sources in both images, we triangulate the position of the 2022 transient, projected on the 2024 image. To address sampling concerns (e.g. Nyquist criteria), we construct an oversampled PSF and fit for the transient's position using a Monte Carlo approach, see Appendix \ref{A:TransientTriangulator} for further discussion\footnote{The open source code to perform these alignments can be found at \href{https://github.com/Astro-Sean/TransientTriangulator/}{https://github.com/Astro-Sean/TransientTriangulator}.}. The results of this triangulation are shown in Figs.~\ref{fig:triangulation} and \ref{fig:contours}. 
	
	We report that both events coincide within $0.02 \pm 0.25$ pixels or $0.007 \pm 0.062$ arcseconds. Given the distance to \host, this corresponds to $2.39 \pm 21.23$ parsecs. To refine this measurement, we conducted systematic tests, including \textit{Jackknife} resampling\footnote{\textit{Jackknife} resampling systematically omits one observation at a time from a dataset, recalculates the statistic of interest for each subset, and aggregates the results to estimate bias and variance \citep{Cochran1946}.}. We measure a refined separation of $2.786 \pm 0.761$ parsecs. This analysis assumes that the \PS/GCP1 CCD remained stable between epochs (we do not account for unexpected image flexure or defects beyond reprojecting each image to correct distortions) and the oversampled PSF adequately fits the source position.
	
	\subsection{Modelling the Progenitor of \mop with \sumo}\label{sec:modelling}
	
	To obtain further insights into the progenitor (system) that lead to \mop, we attempt to model the August 2022 Keck/LRIS spectrum of the first event using the SUpernova MOnte carlo code \citep[\sumo; ][]{Jerkstrand2011, Jerkstrand2012, Jerkstrand2014}. \sumo is a Monte Carlo radiative transfer code operating in non-local thermodynamic equilibrium (NLTE), which solves for the temperature, excitation, and ionisation structure of the ejecta, resulting in a model spectrum. 
	
	\begin{figure*}[!ht]
		\centering
		\includegraphics[width=\textwidth]{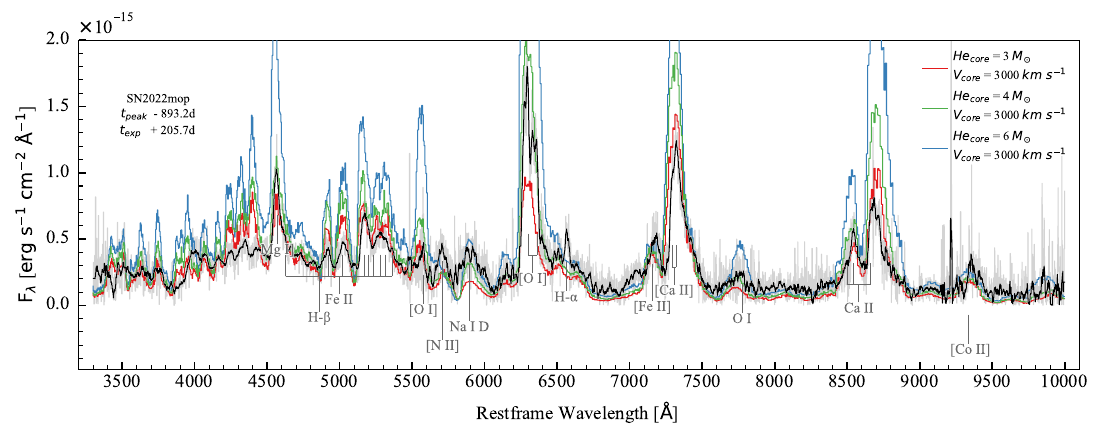 } 
		\caption{A comparison between our \sumo model spectra and the Keck/LRIS spectrum of \mop from August 2022. For clarity, we smooth the observed spectrum (black line, with the original spectrum given in grey). The computed models and \mop spectrum have been scaled to a distance of 10~Mpc, and the models have been simulated with a \Ni mass of 0.06~\msun (which is consistent with SN~1993J, see Fig. \ref{fig:lightcurve}). Line identifications of the most prominent emission lines have been added. The model with a He-core mass of 4~\msun produces the best fit to the observed spectrum.}
		\label{fig:nebular_model_comparison}
	\end{figure*}
	
	Due to the similarities with other SESNe nebular spectra, we use the helium star models from \citet{Woosley2019,Ertl2020} as inputs for spectral modelling. These pre-SN models provide the chemical composition, mass and velocity for SN ejecta for a range of progenitor helium stars of $M_{\text{He, init}}$ $\sim$ 3 -- 12 \msun. Based on comparison to earlier spectra for these models \citep{Dessart2021, Dessart2023,Barmentloo2024}, we create model spectra for three of these progenitors, namely those with $M_{\text{He, init}}$ = 3.3, 4.0 and 6.0 \msun. The final model spectra are presented and compared to \mop in Figure \ref{fig:nebular_model_comparison}. Further details on model setup are given in Appendix \ref{A:sumo_setup}.
	
	The \OIdoublet doublet, which is one of the primary diagnostics for progenitor mass \citep{Jerkstrand2015}, is best reproduced by the \textsc{he4p0} model. The strength of the \NIIdoublet is another diagnostic of progenitor mass \citep{Barmentloo2024}, that may be used for spectra with epochs $\gtrsim$ 200~d. As our estimated epoch for the August 2022 \mop spectrum is 206$\pm$6~d post explosion, we determine the \NIIdiag value as described in \citet{Barmentloo2024} and find \NIIdiag = 6.1, which is equal to the value found for their \textsc{he4p0} model (i.e. $M_{\text{He, init}}$ = 4.0 \msun) at an epoch of 200~d. Another feature of interest in the \mop spectrum is found around 5700\AA. We identify this feature as \NIIsinglet \citep{Jerkstrand2015,Dessart2021,Barmentloo2024}. Comparing to the sample study in \citet{Barmentloo2024}, \NIIsinglet is stronger in \mop than in typical SESNe. Analogously to the \OIsinglet/\OIdoublet pair, \NIIsinglet is a transition from the 2nd to the 1st excited state, whereas \NIIdoublet is from the 1st excited state to the ground state. The presence of \NIIsinglet in the observed spectrum is thus indicative of a relatively high temperature ($T_e \approx 10^4~K$ within the He/N zone) or density, confirming that it must indeed be an early ($\lesssim$ 200 d) nebular spectrum. Furthermore, it confirms the presence of a He/N envelope in \mop, removing the possibility of \mop being a Type Ic event \citep{Barmentloo2024}. 
	
	The nebular modelling points to \mop originating from a star with $M_{\text{He, init}} \sim $ 4.0 \msun. Following the stellar modelling by \citet{Woosley2019,Ertl2020}, such a progenitor would lead to $ M_{\text{ej}} \sim$ 1.6 \msun of hydrogen-poor ejecta. Assuming that the hydrogen envelope was stripped through binary transfer at the start of central helium burning, this translates to $M_{\text{ZAMS}} \sim$ 18 \msun \citep[i.e. following the evolution given in][]{Woosley2019,Ertl2020}. In case the hydrogen envelope was lost later, the extended growth of the He-core through hydrogen shell-burning would require a \textit{lower} $M_{\text{ZAMS}}$ to produce the same mass of hydrogen-poor ejecta and $M_{\text{He, init}}$ mass, meaning we set a limit to the progenitor's $M_{\text{ZAMS}} \lessapprox$ 18 \msun.
	
	\begin{figure}[!ht]
		\centering
		\includegraphics[width=\columnwidth]{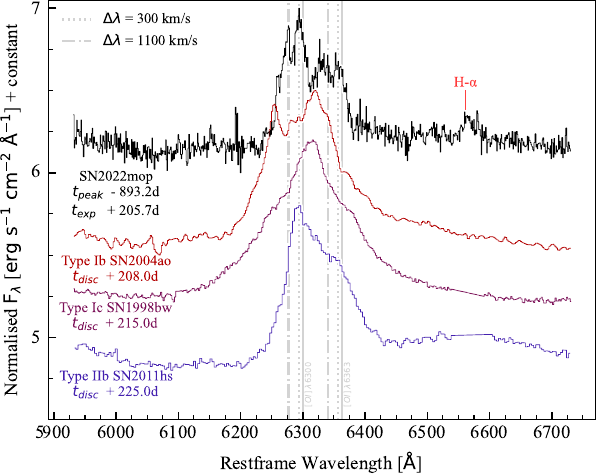 } 
		\caption{Zoom-in of the \OIdoublet profiles for \mop and other stripped-envelope SNe. In \mop, the [O {\sc i}] emission is double-peaked, with the red and blue peaks offset by 300~\kms and 1100~\kms, respectively, from the rest wavelengths.}
		\label{fig:OI_profile}
	\end{figure}
	
	One feature that our models clearly overestimate is the red side of the Ca NIR triplet around 8700 \AA, likely caused by strong [C {\sc ii}] $\lambda$8727 emission. Understanding why this is overproduced is outside of the scope of this work, but will be discussed in a future paper (\textcolor{xlinkcolor}{S. Barmentloo. et al. 2025 in prep}). Due to limitations of the \sumo code, our models do not account for asymmetries in the system. As shown in Fig.~\ref{fig:OI_profile}, the \OIdoublet emission feature exhibits a double-peaked structure, with components blueshifted by 300 and 1100~\kms, respectively. This is a common feature in SESNe nebular spectra \citep{Modjaz2008,Maeda2008,Taubenberger2009,vanBaal2024} and is likely caused by ejecta expanding in a torus- or disk-like geometry.
	
	\subsection{Properties of \mop's Environment}\label{ssec:environment}
	
	\begin{figure}[!ht]
		\centering
		\includegraphics[width=\columnwidth]{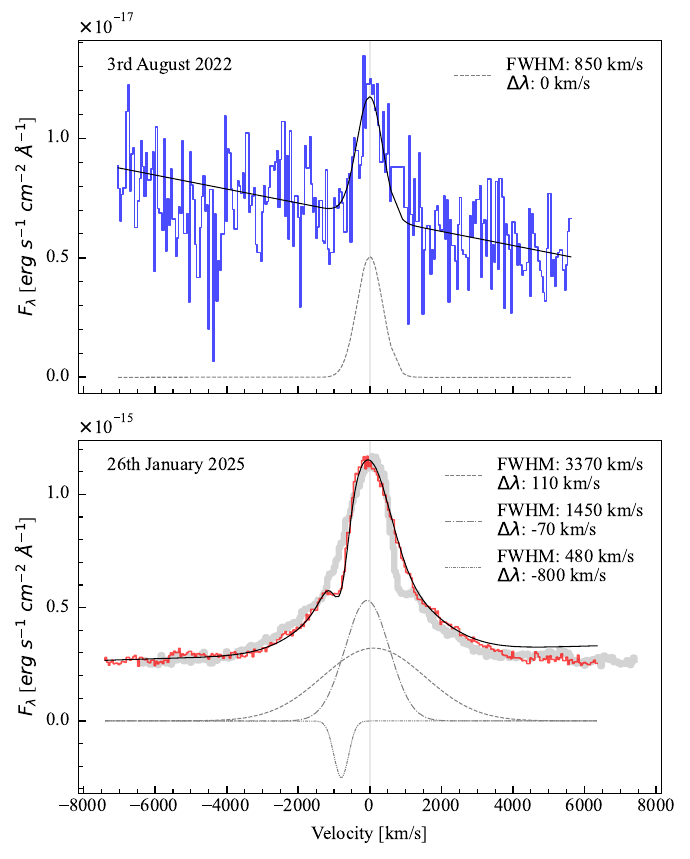 } 
		\caption{Gaussian model fits to the \halpha profile on the 3rd August 2022 (top panel) and 26 January 2025 (bottom panel). The profiles are fitted using a flat continuum over the specified wavelength range. The \halpha profiles show a core FWHM of 850 and 1450~\kms respectively. Note that an unresolved (FWHM = $\sim$135~\kms) noise spike at 6580\AA\ is removed from the 2022 spectrum. The post-peak spectrum shows broad emission with FWHM $\approx 3400$~\kms and a P-Cygni-like feature with an absorption at $-800$~\kms. The thick grey band (lower panel) represents the reflected \halpha profile, highlighting the blueshifted \halpha emission with an extended red excess.}
		\label{fig:FWHM_fits}
	\end{figure}
	
\begin{figure}[!ht]
		\centering
		\includegraphics[width=\columnwidth]{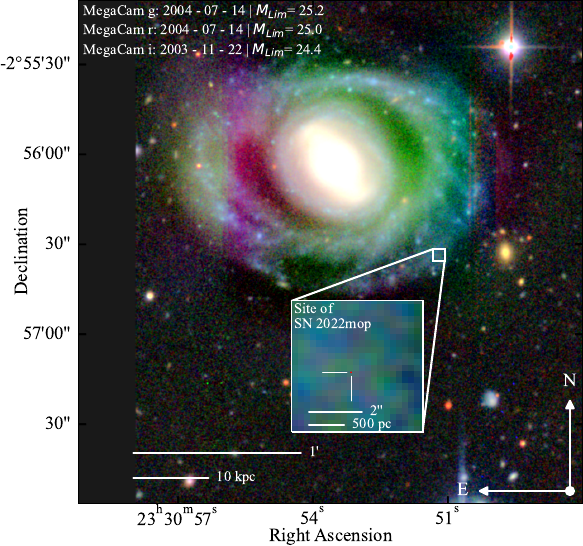} 
		\caption{CFHT/MegaCam $gri$ composite image showing the lenticular host galaxy, \host, and the location of \mop. No bright star-forming cluster is detected within the $1\sigma$ triangulation region, which, (for illustrative purposes) is indicated by the small red circle in the inset.}
		\label{fig:host}
	\end{figure}

	There is significant emission at the position of \halpha in the August 2022 spectrum of \mop (see Fig.~\ref{fig:FWHM_fits}). The width\footnote{This feature is resolved given the Keck/LRIS setup on 3rd August 2022, which used a 1\arcsec slit, with seeing conditions of 1.09\arcsec, and the measured values are FWHM$_{\rm obs} = 18.6$~\AA, FWHM$_{\rm inst} = 6.9$~\AA, with a dispersion of 1.16~\AA/arcsec, yielding a minimum resolution of 320~\kms.} of this feature ($\sim$800~\kms) suggests that it originates from \mop, and any true host emission is likely to be fainter than this. However, we also consider the possibility that it may arise from the underlying host environment. We measure a line luminosity of $1.35 \pm 0.23 \times 10^{-16}~\text{erg}~\text{s}^{-1}~\text{cm}^{-2}$. If this flux originates from the environment surrounding \mop, it would correspond to a star formation rate (SFR) of $\sim 10^{-4}$~\msunperyr using the relation from \citet{Kennicutt1998}. Assuming a flat continuum with an emission profile of similar luminosity, this would correspond to an apparent $r$-band magnitude of $\sim 26$. At this brightness, it would not be detectable in Fig.~\ref{fig:host}, assuming a point-like cluster.
	
	Assuming a Salpeter initial mass function (IMF) \citep{Salpeter1955} of the form $\xi(m) = \xi_{0} \cdot m^{-2.35}$, we estimate the supernova rate (SNR) using the relation $\text{SNR} = \kappa \cdot \text{SFR}$\footnote{For the $SFR_{MW}$ of $\sim 2~M_\odot~{\rm yr}^{-1}$ \citep{Elia2022}, this results in the common consensus of 1--2 Galactic SNe per century \citep{Rozwadowska2021}.}. The efficiency factor ($\kappa$) serves as a proportionality constant and is roughly given as $\kappa \approx \frac{0.01}{~\text{\msunperyr}}$ \citep{Dahlen1999}. Given an SFR of $\sim~10^{-4}$~\msunperyr, the vicinity of \mop would be expected to produce $\sim$1 SN/Myr. 
	
	While the scenario of two isolated transients occurring in the same local environment is disfavoured, it remains difficult to falsify. It is possible that two massive stars originated within the same cluster. Since stars in a cluster are generally assumed to be coeval \citep{Soderblom2010}, these two progenitors must have undergone different evolutionary pathways (but have coordinated core-collapse), as motivated by their distinct chemical signatures. Although one could argue otherwise, given the unique evolution of each transient -- such as the late-time flaring in 2022 and the pre-supernova rise in mid-2024 -- we are motivate to assume that both events occurred within the same binary system and interacted with each other.
	
	\section{Discussion}\label{sec:discussion}
	
	\subsection{\mop and \ip-like transients}\label{sec:09ip}
	
	The eventful history of \mop is reminiscent of \ip\ \citep{Pastorello2013, Fraser2015, Smith2020}, as well as other interacting SNe that exhibit pre-SN activity \citep[see ][]{Ofek2013, Strotjohann2021, Thone2017, Elias-Rosa2016, 2015MNRAS.447.1922M, Fransson2022}. In this area of transient research, pre-SN spectral observations are limited, with the only examples in the literature being \ip \citep[$t_{\text{peak}} - 1122$ days;][]{Smith2010, Foley2011,Pastorello2013}, \bh \citep[$t_{\text{peak}} - 558$ days;][]{Thone2017, Elias-Rosa2016}, and \fyq \citep[$t_{\text{peak}} - 103$ days;][]{Brennan2024a}. For these SNe, the chemical composition of the pre- and post-SN phases is consistent, and the spectral appearance is dominated by signs of ejecta or outflow colliding with the CSM. In the case of \mop, the nebular spectrum taken in 2022 is not consistent with the 2024 observations, suggesting two separate progenitors.
	
	Using epochs with available $ugriz$ bands, we construct a pseudo-bolometric light curve. These epochs include Lulin/SLT observations and span $\pm$10~days post-peak of the 2024 transient. We estimate the blackbody peak luminosity as ${\rm L}_{\mathrm{BB,peak}} \approx 7.8 \times 10^{42}$ erg, with an radiated energy over $\pm 10$ days from peak of ${\rm L}_{\mathrm{int}} \approx 8.9 \times 10^{48}$ erg. This estimate is limited by low-cadence observations and sparse SED coverage but remains consistent with the Event B energetics of \ip \citep[$\sim 10^{49}$ erg; ][]{Smith2014, Moriya2015}. Our first Lulin/SLT observations coincide with the initial rise of the 2024 event. During this period, the blackbody radius ${\rm R}_{\mathrm{BB}}$ expands from $\approx 3.7 \times 10^{14}$ cm at $-10$ d to $\approx 1.1 \times 10^{15}$ cm at $+10$ d, while the blackbody temperature ${\rm R}_{\mathrm{BB}}$ remains roughly constant at $\sim 9000$~K.
	
	The collision of kinematically distinct material is an effective way to convert kinetic energy to radiation, and is typically the dominant power source for interacting SNe. Motivated by this, it has been suggested that several interacting SNe \citep[e.g. ][]{Chugai2004, Dessart2009, Kochanek2012,  Moriya2015} may be non-terminal events, meaning the progenitor star has not undergone core collapse\footnote{While these transients are commonly dubbed ``SN Impostors'' \citep{VanDyk2000,VanDyk2012}, we avoid the use of this nomenclature as it is clear that \mop underwent a genuine core collapse in 2022 due to the strong presence of \OIdoublet (indicative of a large mass of oxygen), the emission from Co [{\sc ii}], and overall nebular appearance.}. We consider the possibility that the 2024 transient is a result late time CSM interaction \citep[e.g. SN 2014C and SN 2001em][]{Margutti2017,Chandra2020}.

	Establishing that the two transients in \mop result from mass loss, core collapse, and \textit{ejecta-shell} interaction is challenging, primarily due to the extreme brightness and energetics of the 2024 transient. The first episode is followed approximately three years later by an even brighter transient  \citep[with energies akin to conventional CCSNe; ][]{Taddia2013,Nyholm2020}, which is difficult to explain outside the context of a pulsational pair-instability supernova \cite[PPISN, which can effectively be ruled out from our August 2022 spectrum; ][]{Woosley2017,Angus2024}. Spectral modelling of the first transient suggests at least $1.6$~\msun of ejecta. Assuming this material is travelling at $\sim$3000~\kms (from our nebular modelling), it is expected to be at $\gtrsim10^{16}$~cm, an order of magnitude greater than the blackbody radius at this time. This argument assumes that the blackbody radius captures the appropriate size of the 2024 event, which may be incorrect if asymmetries or clumping in the CSM is a factor.
	
	Similar events have been observed, showing both spectral metamorphosis and rebrightening at late times \citep[e.g., SNe 2001em and 2014C;][]{Margutti2017, Chandra2020, Thomas2022}. In previous examples, these rebrightening events were mainly observed in radio and X-ray wavelengths, whereas \mop\ exhibits strong optical emission. This difference in late-time evolution is likely an effect of the CSM optical depth. When the optical depth ($\tau$) is greater than $c/v$ (where $v$ is the shock velocity), it primarily produces blackbody optical/UV emission, as observed for \mop. Conversely, when $\tau \ll c/v$, it powers bremsstrahlung X-ray emission as seen for SNe 2001em and 2014C \citep{Margalit2022}.  
	
	If the 2024 transient were influenced by the 2022 ejecta, the second transient should have occurred much earlier. The inconsistency in timing and energetics makes it difficult to reconcile the two events as being related to non-terminal mass-loss activity. Moreover, given that the second peak is brighter, it is reasonable to expect that the second event has greater kinetic energy, likely due to a larger ejecta mass or higher velocity. This raises intriguing questions about how the progenitor system could generate two such energetic eruptions three years apart.

	\begin{figure}[!ht]
		\centering
		\includegraphics[width=\columnwidth]{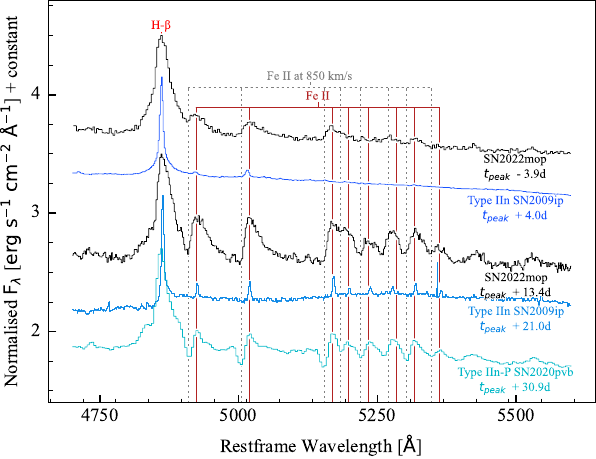 } 
		\caption{The close-up of the approximate $g$-band wavelength range for several interacting transients. We highlight the unique Fe~{\sc ii} emission lines that were not observed for \ip but are seen in several Type IIn SNe, e.g. \pvb \citep{Elias-Rosa2024}. The P Cygni emission for Fe~{\sc ii} lines can be composed of a core emission profile with FWHM = 2000~\kms and an absorption profile at $-$850~\kms, similar to the \halpha profile given in Fig. \ref{fig:FWHM_fits}.}
		\label{fig:FeII_profile}
	\end{figure}
	
	A notable feature of the post-peak spectra of \mop is the strong Fe~{\sc ii} emission, shown in Fig.~\ref{fig:FeII_profile}. This emission resembles a P-Cygni-like profile with absorption at $\sim$800~\kms. Similar profiles have been observed in SNe 2011ht \citep{Mauerhan2012}, 2020pvb \citep{Elias-Rosa2024}, 2016jbu \citep{Kilpatrick2018,Brennan2022c}, and 2015bh \citep{Elias-Rosa2016,Thone2017}. However, such profiles are not present in all \ip-like events, including SNe 2016bdu \citep{Pastorello2018}, and 2009ip itself.
	
	\subsection{A Hydrogen-Poor SN Followed by Mergerburst}\label{sec:mergerburst}
	
	We propose that the transients observed in 2022 and 2024 are physically related and originate from an interacting binary system, ultimately resulting in a ``\textit{mergerburst}'' transient \citep{Chevalier2012,Soker2013,Pastorello2019}\footnote{Or alternatively coined as a Merger-Precursor \citep{Tsuna2024b,Tsuna2024a}.}. Mergerburst transients are driven by the in-spiral and coalescence of massive objects in binary systems with notable examples including V1309 Scorpii \citep{Tylenda2011}, and \etacar \citep{PortegiesZwart2016, Smith2013a, Hirai2021}.
	
	Recent work by \citet{Tsuna2024b,Tsuna2024a} has demonstrated that the precursor emission in \ip-like events, and the smooth rise observed in \fyq \citep{Brennan2024,Dong2024} can be explained by a progenitor star accreting onto a compact object (CO), culminating in a merger event \citep[as suggested for \ip and similar transients by ][and references therein]{Soker2013,Kashi2013,Soker2018}. We demonstrate in Section. \ref{sec:coincidence} that both the 2022 and 2024 events are coincident on the outskirts of \host. In Section. \ref{sec:modelling}, we model the nebular phase spectra and find our best fitting model, \textsc{he4p0}, will result in a neutron star (NS) remnant, \citep[see Figure. 3 in ][]{Ertl2020}. Following the 2022 SN, the newly formed NS may acquire high proper motions due to sudden mass loss and natal kicks \citep{Blaauw1961, Giacobbo2018, Hirai2024}, with typical velocities of 100--300~\kms, occasionally exceeding 1000~\kms \citep{Burrows2023}. If the kick is sufficiently high ($\gtrsim 300~\kms$) and directed appropriately, the NS may interact with its binary companion or its stellar wind/outflows \citep[e.g., SN 2022jli;][]{Moore2023, Chen2024}.
	
	The NS may also receive post-natal slow acceleration that can accelerate it up to $\sim1000$~\kms~$(P_{\rm spin}/1~\mathrm{ms})^{-2}$ ($P_\mathrm{spin}$ is the birth spin period of the NS) on timescales of months to years after the SN, by converting rotational energy into thrust \citep[i.e. the Rocket mechanism; ][]{Harrison1975, Hirai2024}. For a tight mass-transferring binary system, the SN natal kick initially propels the system into a wider and more eccentric orbit. Achieving a direct merger through natal kicks alone is challenging, as it requires fine-tuned conditions within a narrow parameter space. However, once the system is in a wide orbit with a low orbital velocity, the relative influence of the rocket mechanism is enhanced \citep{Hirai2024}. If the rocket effect increases the eccentricity ($e$) sufficiently, approaching $\sim 1$, it can induce a merger. This process occurs with a slightly higher probability, as it requires only two conditions: (a) the natal kick places the system into any sufficiently wide orbit  \citep[e.g. $\sim 30$ days from Fig. \ref{fig:binned_lightcurve}, sufficient to achieve $e$ $\approx$ 1; ][]{Hirai2024}, and (b) the direction of the rocket-induced acceleration is somewhat aligned with the natal kick. The first condition is almost always satisfied due to the Blaauw kick \citep{Blaauw1961} and natal kick, while the second condition may also be relatively common if the suggested correlation between pulsar spin and kick alignment holds \citep[e.g.,][]{Burrows2023}. 
	
	\begin{figure}[!ht]
		\centering
		\includegraphics[width=\columnwidth]{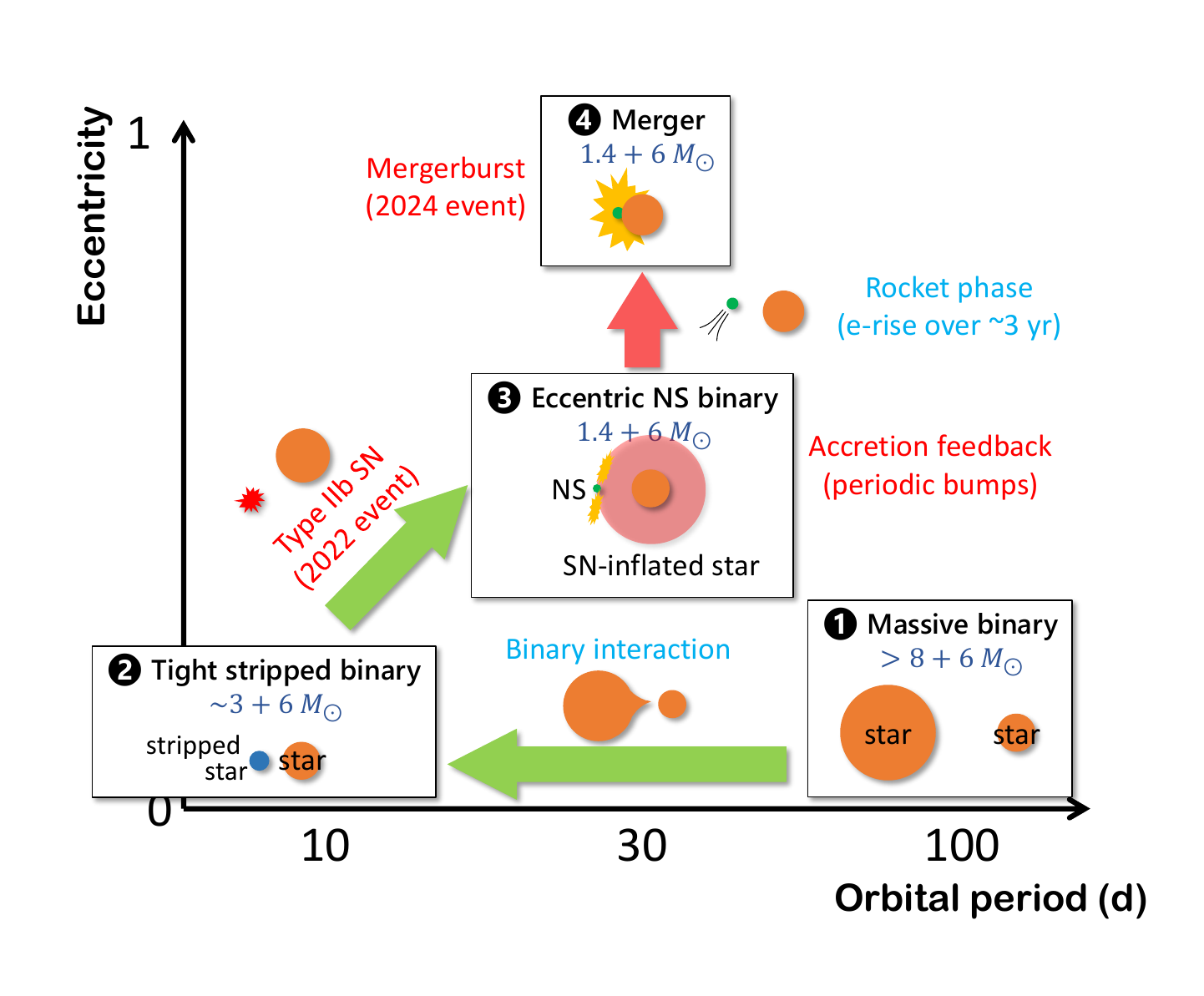 } 
		\caption{Illustration of the proposed mergerburst scenario for \mop.}
		\label{fig:diagram}
	\end{figure}
	
	As the newly formed NS follows an unstable eccentric orbit, it approaches the companion and previously ejected CSM material. Disruption of the outer layers may occur as the NS penetrates the outer envelope. As reported by \citet{Hirai2022b}, these NS-star envelope penetrations would likely be bound, and produce multiple collisions, on the timescales of $\sim$months, similar to that observed for \ip, and in 2022/2023 for \mop (see Fig. \ref{fig:lightcurve}, these flaring events may also be due to periodic feedback from an accreting NS on an eccentric orbit). Due to drag forces within the common envelope, these interactions would subsequently decrease the collision velocity, effectively resulting in a decreased orbital period and cause an in-spiral\footnote{Traditionally, such encounters/collisions were assumed to result in an immediate merger, forming a Thorne–{\text{$\dot{Z}$}}ytkow object \citep[\tzo; ][]{Thorne1975,Hirai2022b} -- a hypothetical system in which a NS core is embedded within a supergiant-like envelope.}. This process transfers orbital energy to the envelope, possibly triggering eruption-like mass loss. The duration of these interactions depends on the initial orbit separation of the NS-star, as well as the magnitude of this drag forces \citep{Hirai2022b}. An illustration of this scenario is given in Fig. \ref{fig:diagram}.
	
	For \ip, no SN-like (i.e. $<$-16~mag) event was observed since first observing the pre-SN progenitor \citep[from HST/WFPC2 F606W in 1999; ][]{Foley2011}, potentially meaning a much larger orbital separation or more eccentric initial orbit i.e. a longer time until complete merger. 
	
	The final outcome of NS-star interaction is unclear. The by-product of this scenario may be a luminous transient powered by super-Eddington accretion onto the engulfed NS \citep{Chevalier1996}, which is likely to be radiatively inefficient and may power large outflows or jets \citep{Moriya2021,Hirai2022b,Soker2022}, possibly producing a ``SN impostor'' event, without disrupting the companion's core. If the NS directly merges with the massive star's core, the combined mass may exceed the \textit{Tolman-Oppenheimer-Volkoff} mass \citep{Oppenheimer1939}, i.e., the maximum mass of a non-rotating neutron star, triggering a violent collapse into a black hole (BH)\footnote{This would be reminiscent of the \textit{collapsar model} \citep[][]{Woosley2006} (however, to date, no gamma-ray burst has been associated with an \ip-like transient, potentially due a result of a \textit{choked-jet} scenario \citep{Senno2016} due the progenitors extended envelope) or the core-merger-induced collapse \citep[CMIC; ][]{Ablimit2022}.} or tidally disrupt the companion's core entirely \citep{Schroder2020}, leading to a merger driven explosion. The formation of a BH may result in significant fallback of nucleosynthesised material, leading to a lack of an obvious nebular phase in \ip-like events, and potentially a large number of Type IIn SNe \citep[e.g. ][]{Fox2013,Fraser2015,Benetti2016,Fox2020}. Such significant fallback may also explain the relatively low-mass ejecta inferred for both Type IIn and Type IIn-p SNe, and also impact the mass of ejected \Ni.
	
	Falsifying this explosion mechanism is difficult, as many of the suspected properties of such an explosion are akin to that of conventional Type IIn SNe \citep{Moriya2021}. Evidence may be found in late-time radio observations, where evidence of a binary induced merger may be collected \citep{Dong2021,Thomas2022}. Alternatively the outcome may result in a transient arising from a common envelope ejection (CCE). This non-terminal explosion would likely leave behind a compact binary system \citep[e.g. SN~2022jli; ][]{Moore2023,Chen2023}. This offers the potential for future deep observations to search for a surviving binary companion \citep[e.g. SN~1993J; ][]{Maund2004,Fox2014}.
	
	\section{Conclusions}\label{sec:conclusions}
	
	We present a $\sim$3-year observation campaign of \mop, a transient that initially exhibited nebular-phase emission in August 2022, resembling a Type Ib/IIb SN at $+$200~d, followed by a Type IIn event almost three years post-explosion in December 2024. We consider the possibility that the 2022 and 2024 events are isolated occurrences that coincidentally took place in the same region on the outskirts of their host galaxy. By triangulating the positions of both transients, we find that they are coincident to within a separation of $\sim$3~pc. Given the low star formation rate in the immediate environment, and the peculiarity of both transients, we are motivated that both events originate from the same locality, likely representing a interacting binary system.
	
	Using the \sumo radiative transfer code, we model the nebular emission and find a relatively unremarkable helium-rich progenitor, likely resulting in a NS remnant from a progenitor with $M_{\text{ZAMS}} \lessapprox$ 18 \msun. We postulate that this remnant enters an eccentric orbit, begins interacting with the companion star. The 2024 event resembles the \textit{Event A/B} light curve seen in the \ip-like class of transients, with a rise time of $\approx 4$~months, reaching $\sim-14$~mag before transitioning into a Type IIn SN, which peaks at $-18.2$~mag. 
	
	\mop represents a compelling example where a compact object can be assumed to be bound to the system, making it a likely contributor to the production of the 2024 Type IIn transient via a merger-burst scenario, involving the newly formed NS and companion star. Continued insight into the possibility of this scenario can be achieved by identifying the pre-SN rise prior to the main SN event, and obtain high cadence photometry, searching for a decaying periodic orbit, a signature of an in-spiralling binary system. Additionally this scenario may emit gravitational waves \citep{Renzo2021}, which may be detectable with the next generation of space-based interferometers \citep[e.g. Laser Interferometer Space Antenna (LISA); ][]{Amaro-Seoane2017} for very nearby transients. However more theoretical work, beyond the scope of this paper, is required to validate this claim. 
	
	Finally, we highlight the uniqueness of \mop.  Without the 2022 Keck/LRIS spectrum presented in this work, \mop would be consistent with the general sample of \ip-like transients and potentially the overall population of interacting SNe. This unprecedented insight highlights the importance of considering binary interaction during the final stages of a massive star’s life. With the additional considerations presented in this work, we stress the necessity for future studies of late-time binary interaction in the context of pre-SN outbursts, \ip-like events, and the broader class of Type IIn SNe. This work is part of a two-part collection focusing on pre-SN variability, interacting SNe, and identification of such activity during the Rubin Era -- see \textcolor{xlinkcolor}{S. J. Brennan et al. 2025, in prep.} for continued discussion.\\
	
	
	\begin{acknowledgements}

S. J. Brennan would like to thank David Jones for their help with the BLAST portal, as well as Takashi Moriya for their insights into merger models. SJB acknowledges their support by the European Research Council (ERC) under the European Union’s Horizon Europe research and innovation programme (grant agreement No. 10104229 - TransPIre). S. S is partially supported by LBNL Subcontract 7707915. SJS acknowledges funding from STFC Grants ST/Y001605/1, ST/X006506/1, ST/T000198/1, a Royal Society Research Professorship and the Hintze Charitable Foundation. AA and T.-W.C acknowledge the Yushan Young Fellow Program by the Ministry of Education, Taiwan for the financial support (MOE-111-YSFMS-0008-001-P1). This publication has made use of data collected at Lulin Observatory, partly supported by MoST grant 109-2112-M-008-001. N. S acknowledges support from the Knut and Alice Wallenberg Foundation through the ``Gravity Meets Light'' project and by and by the research environment grant ``Gravitational Radiation and Electromagnetic Astrophysical Transients'' (GREAT) funded by the Swedish Research Council (VR) under Dnr 2016-06012. W.J-G. is supported by NASA through the NASA Hubble Fellowship grant HSTHF2-51558.001-A awarded by the Space Telescope Science Institute, which is operated by the Association of Universities for Research in Astronomy, Inc., for NASA, under contract NAS5-26555. A.J. acknowledges support by the European Research Council, the Swedish National Research Council, and the Knut and Alice Wallenberg foundation MN is supported by the European Research Council (ERC) under the European Union’s Horizon 2020 research and innovation programme (grant agreement No.~948381) and by UK Space Agency Grant No.~ST/Y000692/1. A. S acknowledges their support from the Knut and Alice Wallenberg foundation through the “Gravity Meets Light” project. M.W.C acknowledges support from the National Science Foundation with grant numbers PHY-2308862 and PHY-2117997. 

		Based on observations made with the Nordic Optical Telescope, owned in collaboration by the University of Turku and Aarhus University, and operated jointly by Aarhus University, the University of Turku, and the University of Oslo, representing Denmark, Finland, and Norway, the University of Iceland, and Stockholm University at the Observatorio del Roque de los Muchachos, La Palma, Spain, of the Instituto de Astrofisica de Canarias. Telescope time on this facility was made available under the Rubin LSST In-Kind program via the SWE-STK-S3 contribution. Pan-STARRS is a project of the Institute for Astronomy of the University of Hawaii, and is supported by the NASA SSO Near Earth Observation Program under grants 80NSSC18K0971, NNX14AM74G, NNX12AR65G, NNX13AQ47G, NNX08AR22G, 80NSSC21K1572, and by the State of Hawaii. Based on observations obtained with the Samuel Oschin 48-inch Telescope and the 60-inch Telescope at the Palomar Observatory as part of the Zwicky Transient Facility (ZTF) project. ZTF is supported by the National Science Foundation under Grant No. AST-2034437 and Award No. 2407588, as well as a collaboration including Caltech, IPAC, the Weizmann Institute of Science, the Oskar Klein Center at Stockholm University, the University of Maryland, Deutsches Elektronen-Synchrotron, Humboldt University, the TANGO Consortium of Taiwan, the University of Wisconsin at Milwaukee, Trinity College Dublin, Lawrence Livermore National Laboratories, IN2P3, the University of Warwick, Ruhr University Bochum, Cornell University, Northwestern University, Drexel University, the University of North Carolina at Chapel Hill, and the Institute of Science and Technology, Austria. Operations are conducted by Caltech's Optical Observatory (COO), Caltech/IPAC, and the University of Washington at Seattle. The Gordon and Betty Moore Foundation, through both the Data-Driven Investigator Program and a dedicated grant, provided critical funding for SkyPortal. Funded by the European Union (ERC, project number 101042299, TransPIre). Views and opinions expressed are however those of the author(s) only and do not necessarily reflect those of the European Union or the European Research Council Executive Agency. Neither the European Union nor the granting authority can be held responsible for them.
	\end{acknowledgements}
	
	
	
	\bibliographystyle{aa}
	\bibliography{references.bib}
	
	\begin{appendix}
		
		\section{Observations logs and Reductions}\label{A:obs_log}
		
		\subsection{Photometric Reductions}
		
		\begin{figure}[!ht]
			\centering
			\includegraphics[width=\columnwidth]{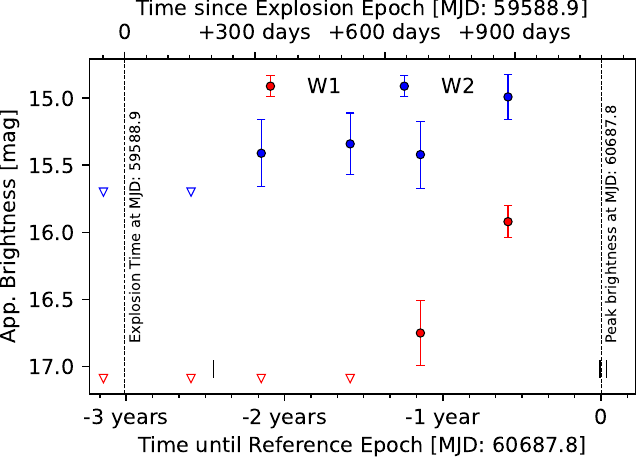 } 
			\caption{WISE Lightcurve of \mop with coverage extending before the 2022 transient of \mop as well as the initial rise before the 2024 event. While this may represent a late time rebrightening at longer wavelengths due to nebular emission from \mop, it is possible this reflects ongoing CSM interaction or dust formation.}
			\label{fig:wise_lightcurve}
		\end{figure}
		
		The \autophot\ pipeline \citep{Brennan2022d} was used to perform photometric measurements when a dedicated photometry service was unavailable, as well as to visually confirm pre-SN activity in the ZTF images. For each dataset, the target's coordinates were obtained from the Transient Name Server \citep[TNS; ][]{Gal-Yam2021}, and the World Coordinate System (WCS) values of the images were verified using \texttt{Astrometry.net} \citep{Lang2010}. An effective point spread function (ePSF) model was constructed using bright, isolated sources in the image with the \texttt{Photutils} package \citep{bradley_2024}. The PSF was fitted using a Markov Chain Monte Carlo (MCMC) approach, implemented in a custom MCMC fitter class within \texttt{LMFIT} \citep{Newville2024} and \texttt{emcee} \citep{Foreman-Mackey2013}. This class defines the prior and likelihood functions based on the model parameters and incorporates the noise variance in the fitting procedure. The MCMC sampler was used to explore the parameter space and obtain the best-fit model parameters for position and flux, including uncertainties.  The $griz$ band zero-points were calibrated against sequence sources in the ATLAS All-Sky Stellar Reference Catalogue \citep[REFCAT2;][]{Tonry2018} and the Sloan Digital Sky Survey \citep[SDSS; ][]{Abazajian2009} for $u$-band. Science and reference images were aligned using \texttt{Astropy}'s \citep{astropy:2013,astropy:2018,astropy:2022} \texttt{REPROJECT} function, and reference image subtraction was performed using the saccadic fast Fourier transform (\texttt{SFFT}) algorithm \citep{Hu2022} to isolate any potential transient flux. If transient flux was not detected at a 5$\sigma$ confidence level, the limiting magnitude of the image was determined via artificial source injection and recovery, as detailed in \citet{Brennan2022d}.

		\begin{table*}[!ht]
\centering\caption{Photometric log for \mop. The phase of each observation is given as $t_{\text{exp}}$ = 59588.9 and $t_{\text{peak}}$ = 60687.8. Full table available in online supplementary material.}
\label{tab:SN2022mop_photometry_log}
\begin{tabular}{lccccccc}
\hline
Date & MJD & $t - t_{peak}$ (days) & $t - t_{exp}$ (days) & Magnitude & Error & Filter & ID \\ \hline
\midrule
2022-01-13 & 59592.24 & -1095.5 & +3.3 & 18.832 & 0.144 & o & ATLAS \\
2022-01-21 & 59600.23 & -1087.5 & +11.3 & 18.793 & 0.157 & o & ATLAS \\
2022-06-02 & 59732.58 & -955.2 & +143.7 & 20.207 & 0.190 & c & ATLAS \\
2022-06-06 & 59736.59 & -951.2 & +147.7 & 20.216 & 0.196 & c & ATLAS \\
2022-06-10 & 59740.43 & -947.4 & +151.5 & 20.764 & 0.044 & g & ZTF \\
2022-06-10 & 59740.46 & -947.3 & +151.6 & 20.610 & 0.031 & r & ZTF \\
2022-06-11 & 59741.58 & -946.2 & +152.7 & 20.448 & 0.105 & w & Pan-Starrs \\
\bottomrule
\end{tabular}
\end{table*}

		\begin{table*}[!ht]
\centering\caption{
Photometric log for SN 2022mop for WISE W1 and W2 bands. The phase of each observation is given as 
$t_{\text{exp}}$ = 59588.9 and $t_{\text{peak}}$ = 60687.8. 
We perform photometry on the subtracted images using a standard 8.25\arcsec\ radius aperture, applying suitable aperture corrections and zero magnitude fluxes for the WISE W1 and W2 bands 
(with central wavelengths of 3.4 and 4.6 µm, respectively) of 
$F_{\nu,0}^{W1}$ = 306.7 Jy and $F_{\nu,0}^{W2}$ = 170.7 Jy \citep{Wright2010}. 
Full table available in online supplementary material.
}
\label{tab:SN2022mop_WISE_photometry_log}
\begin{tabular}{lcccccccc}
\hline
Date & MJD & $t - t_{peak}$ (days) & $t - t_{exp}$ (days) & $\mu$Jy & $\Delta\mu$Jy & Magnitude & Error & Filter \\ \hline
\midrule
2016-06-12 & 57551.00 & -3136.8 & -2037.9 & $>$45.0 & - & $>$17.09 & - & W1 \\
2016-06-12 & 57551.00 & -3136.8 & -2037.9 & $>$90.0 & - & $>$15.7 & - & W2 \\
2016-11-28 & 57720.00 & -2967.8 & -1868.9 & $>$45.0 & - & $>$17.09 & - & W1 \\
2016-11-28 & 57720.00 & -2967.8 & -1868.9 & $>$90.0 & - & $>$15.7 & - & W2 \\
2017-06-11 & 57915.00 & -2772.8 & -1673.9 & $>$90.0 & - & $>$15.7 & - & W2 \\
\bottomrule
\end{tabular}
\end{table*}

		\subsection{Spectroscopy}
		
\begin{figure*}[!ht]
			\centering
			\includegraphics[width=\textwidth]{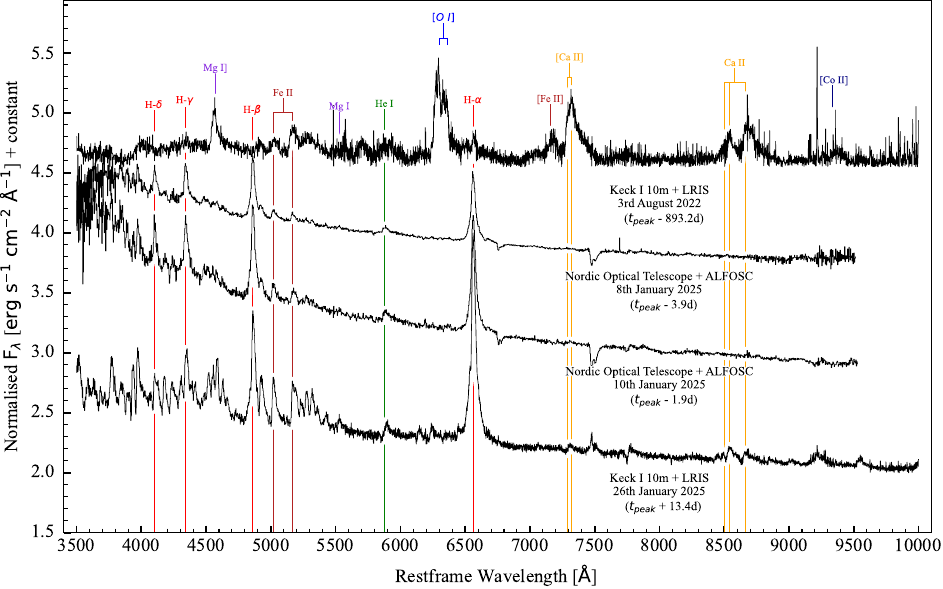 } 
			\caption{Spectroscopic observations for \mop. The phase of each observation is given as $t_{\text{peak}}$ = 60687.8.}
			\label{fig:all_spectra}
		\end{figure*}

		The complete spectroscopic dataset of \mop is presented in Fig. \ref{fig:all_spectra} with an observational log given in Table. \ref{tab:SN2022mop_spectroscopy_log}. Calibrated spectra available on WISeREP \citep{Yaron2012}.

		\begin{table*}[!ht]
\centering\caption{ Spectropscopic log for \mop. The phase of each observation is given as $t_{\text{exp}}$ = 59588.9 and $t_{\text{peak}}$ = 60687.8. All spectra observed \mop using a 1$\arcsec$ slit. Calibrated spectra available on WISeREP \citep{Yaron2012}. }
\label{tab:SN2022mop_spectroscopy_log}
\begin{tabular}{cccccccc}
\hline
Date & MJD & $t - t_{peak}$ (days) & $t - t_{exp}$ (days) & Telescope & Instrument & Grating & Exposure (s) \\ \hline
\midrule
2022-08-03 & 59794.6 & -893.2 & +205.7 & Keck I 10m & LRIS & 400/8500 & 900 \\
2025-01-08 & 60683.9 & -3.9 & +1095.0 & Nordic Optical Telescope & ALFOSC & Grism \#4 & 1200 \\
2025-01-10 & 60685.8 & -1.9 & +1096.9 & Nordic Optical Telescope & ALFOSC & Grism \#4 & 1200 \\
2025-01-26 & 60701.2 & +13.4 & +1112.3 & Keck I 10m & LRIS & 400/8500 & 900 \\
\bottomrule
\end{tabular}
\end{table*}

		\section{Estimating the explosion time of \mop}\label{A:explosion_epoch}
		
\begin{figure}[!ht]
			\centering
			\includegraphics[width=\columnwidth]{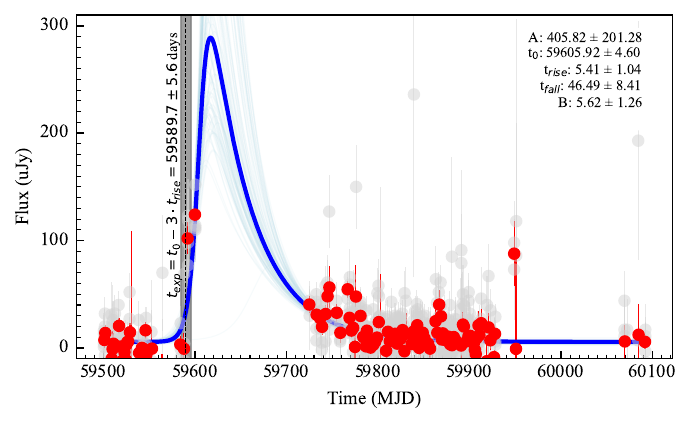 } 
			\caption{Estimate of the explosion time of \mop using a Bazin function fit. Grey points represent the raw data, while red points correspond to 1-day binned data. The blue lines show fits obtained using \texttt{emcee} sampling with the \texttt{LMFIT} package.}
			\label{fig:explosion_time}
		\end{figure}

		We estimate the explosion time of \mop using the Bazin function \citep{Bazin2009} based on flux measurements from the ATLAS forced photometry server. We estimate the explosion epoch ($t_{\text{exp}}$) as the peak ($t_0$) minus 3 times the rise time ($t_{\text{rise}}$) and report a value of $t_{\text{exp}} = 59589.9 \pm 5.5$ (around January 2022), see Fig. \ref{fig:explosion_time}. Given the low photometric cadence, this is a rough approximation to the explosion epoch.
		
		\section{\sumo setup}\label{A:sumo_setup}
		
		When preparing these models for \sumo, the ejecta are divided into seven different compositional zones. Named after their dominant elements, these are the Fe/He, Si/S, O/Si/S, O/Ne/Mg, O/C, He/C, and He/N zones. Of these zones, the first five are assumed to be fully macroscopically mixed, meaning that the zones are split into clumps of distinct composition. These clumps are then mixed in velocity space but retain their distinct composition (opposite to microscopic mixing, where all zones simply diffuse into a single ``soup'' of elements). We also mix in 40\%, 18\%, and 10\% of the He/C zone into this core region for the $M_{\text{He, init}} = 3.3$, 4.0, and 6.0 \msun models, respectively. This macroscopic mixing thus leads to the line widths and structures of most spectral lines in the nebular phase being very similar. From measurements of some of the lines in the \mop spectrum, we find that this macroscopically mixed core is contained within $V_{\text{core}} \sim 3000$ \kms. We use this observation as a constraint in our spectral modelling by forcing the first five zones to be contained within 3000 \kms. Furthermore, we artificially increase the volume of the Fe/He and Si/S zones, as these zones are predicted to expand in the first few days after the SN due to radioactive heating, following the treatment described in Section 3.2.1 in \citet{Jerkstrand2015}. Another change we make compared to the models by \citet{Woosley2019} and \citet{Ertl2020} is the \Ni mass. The \Ni masses found in the original models ($\sim 0.02 - 0.04$ \msun) are significantly lower than what is found in observations, roughly 0.07 -- 0.15 \msun \citep{Prentice2016, Taddia2018, Prentice2019,Rodriguez2023}. We therefore increase this mass by artificially increasing the mass of the Fe/He zone in the model, leaving its composition unchanged. The late-time lightcurve curve of \mop does not allow for an accurate \Ni estimate. Based on matching to the quasi-continuum flux in the observed spectrum, we set the \Ni mass in each model to 0.06~\msun. 
		
		\section{Triangulation of the 2024 transient}\label{A:TransientTriangulator}
		
		Data from \PS provide images for both epochs, with a pixel scale of 0.258 arcsec/pixel, corresponding to a projected 88 parsec/pixel at the distance of \host. We select epochs in September 2022 and October 2024 when the respective transients were at the approximate same magnitude. To triangulate the position of \mop, we build a set of 3-point asterisms using common sources in the 2022 image, employing two bright sources as anchor points, and the position of \mop as the third vertex. An oversampled ePSF is constructed using the \autophot code. Sources are excluded if their GAIA parallax exceeds 1 mili-arcsecond (which is smaller than the typical fitting errors of each source), or if their PSF profile is extended (i.e., saturated, diffuse sources) in the either image. Using these pre-built asterisms, we project the expected 2022 position of \mop onto the 2024 \PS images by using the same anchor sources and projecting the expected location on the third vertex. To account for position fitting error, we perform bootstrapped measurements by applying a pixel shift to each anchor position. These shifts follow a normal distribution with a standard deviation corresponding to the fitting error
		

	\end{appendix}
\end{document}